\documentclass[12pt]{article}\usepackage[]{graphicx}\usepackage[]{color}
\makeatletter
\def\maxwidth{ %
  \ifdim\Gin@nat@width>\linewidth
    \linewidth
  \else
    \Gin@nat@width
  \fi
}
\makeatother

\definecolor{fgcolor}{rgb}{0.345, 0.345, 0.345}
\newcommand{\hlnum}[1]{\textcolor[rgb]{0.686,0.059,0.569}{#1}}%
\newcommand{\hlstr}[1]{\textcolor[rgb]{0.192,0.494,0.8}{#1}}%
\newcommand{\hlopt}[1]{\textcolor[rgb]{0,0,0}{#1}}%
\newcommand{\hlstd}[1]{\textcolor[rgb]{0.345,0.345,0.345}{#1}}%
\newcommand{\hlkwa}[1]{\textcolor[rgb]{0.161,0.373,0.58}{\textbf{#1}}}%
\newcommand{\hlkwb}[1]{\textcolor[rgb]{0.69,0.353,0.396}{#1}}%
\newcommand{\hlkwc}[1]{\textcolor[rgb]{0.333,0.667,0.333}{#1}}%
\newcommand{\hlkwd}[1]{\textcolor[rgb]{0.737,0.353,0.396}{\textbf{#1}}}%

\usepackage{framed}
\makeatletter
\newenvironment{kframe}{%
 \def\at@end@of@kframe{}%
 \ifinner\ifhmode%
  \def\at@end@of@kframe{\end{minipage}}%
  \begin{minipage}{\columnwidth}%
 \fi\fi%
 \def\FrameCommand##1{\hskip\@totalleftmargin \hskip-\fboxsep
 \colorbox{shadecolor}{##1}\hskip-\fboxsep
     \hskip-\linewidth \hskip-\@totalleftmargin \hskip\columnwidth}%
 \MakeFramed {\advance\hsize-\width
   \@totalleftmargin\z@ \linewidth\hsize
   \@setminipage}}%
 {\par\unskip\endMakeFramed%
 \at@end@of@kframe}
\makeatother

\definecolor{shadecolor}{rgb}{.97, .97, .97}
\definecolor{messagecolor}{rgb}{0, 0, 0}
\definecolor{warningcolor}{rgb}{1, 0, 1}
\definecolor{errorcolor}{rgb}{1, 0, 0}
\newenvironment{knitrout}{}{} 

\usepackage{alltt} 
\usepackage{natbib}
\usepackage{amsmath}
\usepackage{amssymb}
\usepackage{fullpage}

\newif\ifblinded
 \blindedfalse 
\ifblinded
{ }
\else
 \author{E. L. Ionides, C. Breto, J. Park, R. A. Smith and A. A. King\\
 University of Michigan
 }
\fi

\title{Monte Carlo profile confidence intervals}
\IfFileExists{upquote.sty}{\usepackage{upquote}}{}
\begin{document}

\newcommand\bias{\beta}

\newcommand\prob{\mathbb{P}}
\newcommand\E{\mathbb{E}}
\newcommand\var{\mathrm{Var}}
\newcommand\cov{\mathrm{Cov}}
\newcommand\loglik{\ell}
\newcommand\R{\mathbb{R}}
\newcommand\data[1]{#1^*}
\newcommand\param{\, ; \,}
\newcommand\transpose{\scriptsize{T}}
\newcommand\lik{\mathscr{L}}
\newcommand\logLik{\ell} 
\newcommand\profileLogLik[1]{\ell^\mathrm{profile}_#1}
\newcommand\given{{\, | \,}}
\newcommand\SE{\mathrm{SE}}
\newcommand\dimTheta{p}
\newcommand\genPomp{\texttt{genPomp}}
\newcommand\mycolon{\,{:}\,}
\newcommand\Xspace{\mathbb{X}}
\newcommand\Yspace{\mathbb{Y}}


\maketitle

\begin{abstract}
Monte Carlo methods to evaluate and maximize the likelihood function enable the construction of confidence intervals and hypothesis tests, facilitating scientific investigation using models for which the likelihood function is intractable.
When Monte Carlo error can be made small, by sufficiently exhaustive computation, then the standard theory and practice of likelihood-based inference applies.
As data become larger, and models more complex, situations arise where no reasonable amount of computation can render Monte Carlo error negligible.
We develop profile likelihood methodology to provide frequentist inferences that take into account Monte Carlo uncertainty.
We investigate the role of this methodology in facilitating inference for computationally challenging dynamic latent variable models.
We present three examples arising in the study of infectious disease transmission. 
These three examples demonstrate our methodology for inference on nonlinear dynamic models using genetic sequence data, panel time series data, and spatiotemporal data.
We also discuss applicability to nonlinear time series analysis.
\end{abstract}

\vspace{1mm}

\begin{center}
\end{center}

\section{Introduction}
 
This paper develops profile likelihood inference methodology for situations where computationally intensive Monte Carlo methods are employed to evaluate and maximize the likelihood function.
If the profile log likelihood function can be computed with a Monte Carlo error small compared to one unit, carrying out statistical inference from the Monte Carlo profile as if it were the true profile will have relatively small effects on resulting confidence intervals. 
Sometimes, no reasonable amount of computation can reduce the Monte Carlo error in evaluating the profile to levels at or below one log unit. 
This predicament typically arises with large datasets and complex models.
However, to investigate large datasets in the context of complex models there is little alternative to the use of Monte Carlo methods.
We develop an approach to effective likelihood-based statistical inference taking into account the non-negligible Monte Carlo error.
We choose to focus on Monte Carlo profile likelihood confidence intervals, since their construction gives convenient opportunities to assess  Monte Carlo variability and make appropriate compensations.

Our paper is organized as follows. 
First, we set up mathematical notation to formalize the task of Monte Carlo profile likelihood estimation via a metamodel. 
Section~\ref{sec:related} puts this task in the context of some previous work on likelihood-based inference for intractable models.
Section~\ref{sec:cutoff} develops our methodological approach.
Section~\ref{sec:pomp} presents a dynamic latent variable modeling framework of broad applicability for which the methodology is appropriate. 
For this class of models, we demonstrate the capabilities of our methodology by solving three inferential challenges, each representing a different data type for which scientific progress is limited by the availability of effective statistical methodology.
These examples all arise from the study of transmissible human diseases, a field characterized by extensive and diverse data, indirectly observation of the underlying infection processes, strongly nonlinear stochastic dynamics, and public health importance.
Infectious disease data therefore provide many inference opportunities and challenges.
Section~\ref{sec:genPomp} concerns inference on population dynamics from genetic data; Section~\ref{sec:panelPomp} concerns  fitting nonlinear partially observed Markov models to panel data; Section~\ref{sec:spacetime} concerns fitting a nonlinear partially observed spatiotemporal model.
Section~\ref{sec:timeseries} discusses the role of our methodology in nonlinear time series analysis.
Section~\ref{sec:sim} investigates our methodology via a simulation study on a toy example.
Section~\ref{sec:discussion} is a concluding discussion which situates our paper within the broader goal of inference for large datasets and complex models.

\begin{figure}
\begin{knitrout}\small
\definecolor{shadecolor}{rgb}{0.969, 0.969, 0.969}\color{fgcolor}

{\centering \includegraphics[width=\maxwidth]{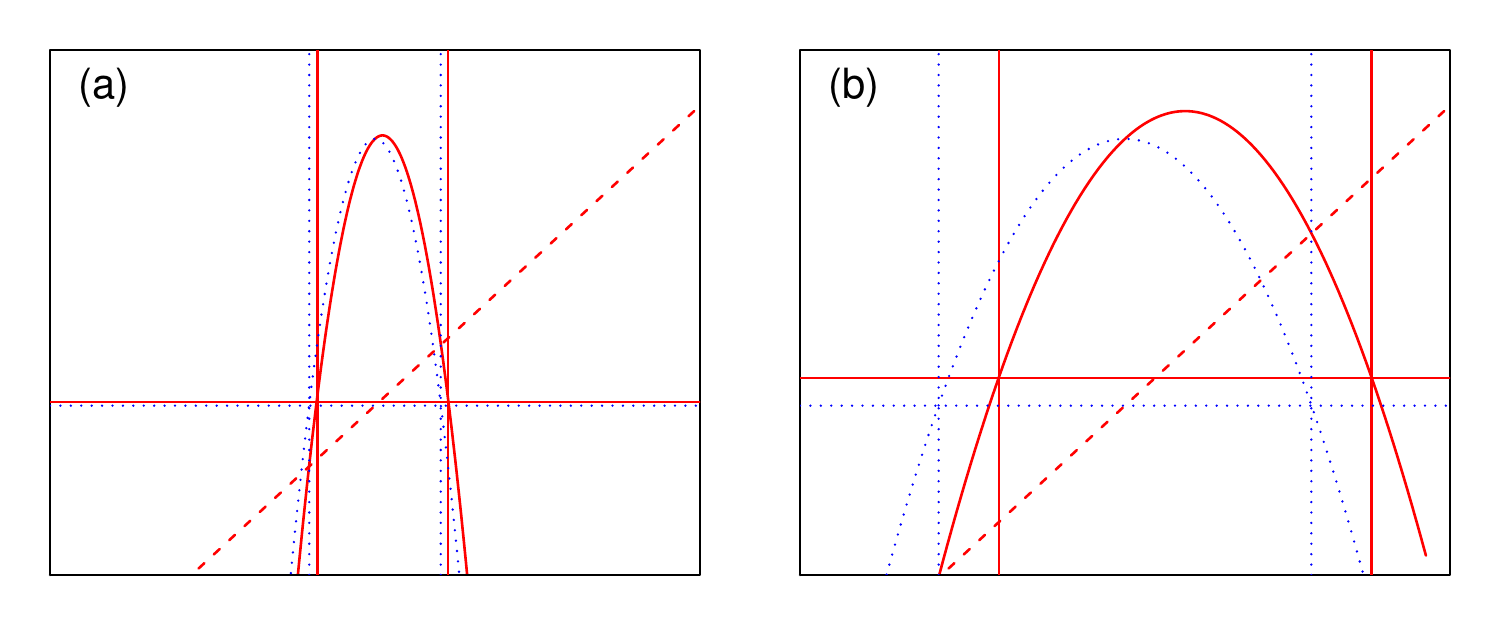} 

}

\end{knitrout}
\caption{The effect of bias on profile likelihood confidence intervals, demonstrated on a quadratic profile log likelihood function.
The two examples represent: (a) high information about the profiled parameter, i.e., high curvature of the profile log likelihood relative to the slope of the bias; (b) lower information.
A quadratic profile log likelihood and its corresponding confidence interval are shown as blue dotted lines.
Linear bias, $\beta_k=c_0+c_1k$, is shown as a red dashed line.
A profile including this bias, and the corresponding biased confidence intervals, are shown as red solid lines.
}\label{fig:bias}
\end{figure}

We consider a general statistical inference framework in which data are a real-valued vector, $\data{y}$, modeled as a realization of a random variable $Y$ having density $f_Y(y\param\theta)$, where $\theta$ is an unknown parameter in $\R^{\dimTheta}$.
We are concerned with inference on $\theta$ in situations where the data analyst cannot directly evaluate $f_Y(y\param\theta)$.
Instead, we suppose that approximate evaluation of  $f_Y(y\param\theta)$ is possible through Monte Carlo approaches. 
One situation in which this arises is when the statistician can simulate draws from the density $f_Y(y\param\theta)$ despite being unable to directly evaluate it \citep{diggle84}.
In addition to a simulator for the full joint distribution of $Y$, we might also have access to simulators for various marginal and conditional distributions related to $f_Y(y\param\theta)$.
For example, this can arise if $Y$ has the structure of a fully or partially observed Markov process \citep{breto09}.
Simulation-based methods are growing in usage, motivated by advances in the availability of complex data and the desire for statistical fitting of complex models to these data. 
Although we cannot calculate them, we can nevertheless define the log likelihood function,
\begin{equation} \label{LogLik} 
\loglik(\theta\param\data{y})=\log f_Y(\data{y}\param\theta),
\end{equation}
and a maximum likelihood estimate (MLE),
\begin{equation} \label{MLE} 
\hat\theta^* = \hat\theta(\data{y}) = \arg \max_\theta \loglik(\theta\param \data{y}).
\end{equation}
To formalize the task of constructing marginal confidence intervals, we suppose that $\theta=(\phi,\psi)$ with $\phi\in\R^1$ and  $\psi\in\R^{\dimTheta-1}$. 
Here, $\phi$ is a focal parameter for which we are interested in obtaining a confidence interval using the data, $\data{y}$.
By changing the focal parameter, we are equivalently interested in the general problem of obtaining a marginal confidence interval for each component of a parameter vector. 
The profile log likelihood function for $\phi$ is defined as
\begin{equation} \label{Profile} 
\loglik^P(\phi\param\data{y})=\max_\psi \loglik\big((\phi,\psi)\param\data{y}\big).
\end{equation}
The profile log likelihood is maximized at a marginal MLE, 
\begin{equation}\label{marginal_mle}
\hat\phi^*=\hat\phi(\data{y})=\arg\max_\phi \loglik^P(\phi\param\data{y}).
\end{equation}
A profile likelihood confidence interval with cutoff $\delta$ is defined as
\begin{equation} \label{def_profile_ci}
\big\{\phi:  \loglik^P(\phi\param\data{y}) > \loglik^P\big(\hat\phi^*\param\data{y}\big) - \delta \big\}.
\end{equation}
Profile likelihood confidence intervals are a widespread inference approach with some favorable properties, including asymptotic efficiency and natural transformation under reparameterization \citep{pawitan01}.
Modifications can lead to higher-order asymptotic performance \citep{barndorff94} but these are not routinely available.
In our context \eqref{Profile}, \eqref{marginal_mle} and \eqref{def_profile_ci} are not directly accessible to the data analyst.
Instead, we work with independent Monte Carlo profile likelihood evaluations at a sequence of points $\phi_{1:K}=(\phi_1,\dots,\phi_K)$. 
We denote the evaluations as $\big(\breve \loglik^P_k(\data{y}),k\in 1{\mycolon}K\big)$, using a breve accent to distinguish Monte Carlo quantities. 
We write a decomposition,
\begin{equation} \label{ProfileMetamodel}
\breve \loglik^P_k(\data{y}) = \loglik^P(\phi_k\param\data{y})+ \bias_k(\data{y}) + \epsilon_k(\data{y}), \quad k\in 1{\mycolon}K,
\end{equation}
where $\epsilon_{1:K}(Y)$ are Monte Carlo random variables which are, by construction, mean zero and independent conditional on $Y$. 
Thus, $\bias_{1:K}(\data{y})$ gives the Monte Carlo bias of each profile log likelihood evaluation.
If the amount of information about $\phi$ in the data is large, the curvature of the profile log likelihood is large and the statistically relevant region of high likelihood is narrow.
In that case, it may be reasonable to approximate the Monte Carlo bias and error distribution as constant across the profile, modeling $\epsilon_{1:K}(\data{y})$ as conditionally independent and identically distributed with. 
We suppose that the conditional variance of $\epsilon_k$ is $\sigma^2(\data{y})<\infty$, and we write $\epsilon_k\sim\mathrm{IID}\big(\sigma^2(\data{y})\big)$.
The metamodel becomes,
\begin{equation} \label{ProfileMetamodelUnbiased}
\breve \loglik^P_k(\data{y}) = \loglik^P(\phi_k\param\data{y})+ \bias(\data{y}) + \epsilon_k, \quad \epsilon_k\sim \mathrm{IID}\big(\sigma^2(\data{y})\big), \quad k\in 1{\mycolon}K.
\end{equation}
Emprical evidence for non-constant Monte Carlo variance could motivate the inclusion of heteroskadistic errors in \eqref{ProfileMetamodelUnbiased}.
The assumption in \eqref{ProfileMetamodelUnbiased} of approximately constant Monte Carlo bias is hard to quantify empirically on challenging computational problems, since one cannot readily obtain an estimate with negligible bias.
Figure~\ref{fig:bias} demonstrates pictorially the consequence of linear bias on confidence intervals constructed for a quadratic profile log likelihood function.
We see that linear bias, $\beta_k=c_0+c_1k$, affects the location of the maximum of the profile but does not affect the curvature and therefore has no effect on the width of the resulting profile confidence interval.
Although the bias on the Monte Carlo profile likelihood estimate may be intractable, the coverage of a constructed confidence interval can be checked by simulation at a specific parameter value such as an MLE, as demonstrated in Section~\ref{sec:sim}.

\section{Previous work on likelihood-based inference via simulation}\label{sec:related}

A prescient paper by \citet{diggle84} developed Monte Carlo maximum likelihood methodology with similar motivation to our current goals.
However, \citet{diggle84} did not work with profile likelihood and did not show how to correct the resulting confidence intervals for Monte Carlo error.
Further, \citet{diggle84} assumed that the Monte Carlo methods would involve simulating from the modeled joint distribution of the entire dataset, whereas modern computationally efficient Monte Carlo algorithms may be based on simulating sequentially from conditional distributions in a carefully crafted decomposition of the entire joint distribution. 
\citet{breto09} and  \citet{he10} introduced the term {\it plug-and-play} to describe statistical methodology for which the model (viewed as an input to an inference algorithm) is specified via a simulator in this broader sense.
The term {\it likelihood-free} has been used similarly, in the context of Markov chain Monte Carlo \citep{marjoram03} and sequential Monte Carlo \citep{sisson07}.
The term {\it equation-free} has been used for the related concept of simulation-based model investigations in the physical sciences \citep{kevrekidis04}.
Related terms {\it implicit} \citep{diggle84} and {\it doubly intractable} \citep{lyne15} have been used to describe models for which only plug-and-play algorithms are practical.
From the point of view of categorizing statistical methodology, it is convenient to view the way in which an inference algorithm accesses the statistical model as a property of the algorithm rather than a property of the model. 

\citet{rubio13} investigated nonparametric estimation of a likelihood surface via approximate Bayesian computing (ABC). 
\citet{rubio13} also provided a literature review of previous approaches to carry out statistical inferences in situations where likelihood evaluation and maximization necessarily involve computationally intensive and noisy Monte Carlo procedures.
We are not aware of previous work developing Monte Carlo profile likelihood methodology.
Profile methodology focuses the computational effort on parameters of key interest---specifically, those for which one computes the profile.
The process of constructing a profile requires computation of a relevant feature of the likelihood surface in the region of inferential interest.
Studying the likelihood surface on this scale, rather than focusing exclusively on a point estimate such as the maximum likelihood estimate, has some theoretical justification \citep{ionides05}.
In the general theory of stochastic simulation-based optimization, building metamodels describing the response surface is a standard technique \citep{barton06}.
Our goal is to develop metamodel methodology that takes advantage of the statistical properties of the profile likelihood and constructs confidence intervals correcting properly for Monte Carlo variability.

\section{Profile cutoff correction via a local quadratic metamodel}\label{sec:cutoff}

Local asymptotic normality (LAN) provides a general theoretical framework in which the log likelihood function is asymptotically well approximated by a quadratic \citep{lecam00}.
Under sufficient regularity conditions, this quadratic approximation is inherited by the profile log likelihood \citep{murphy00}.
Here, we write the marginal $\phi$ component of the LAN property as a finite sample normal approximation given by
\begin{equation}\label{LAN} 
\loglik^P(\phi\param Y) - \loglik^P(\phi_0\param Y) \approx Z(\phi-\phi_0)\sqrt{I}-(\phi-\phi_0)^2I/2,
\end{equation}
where $Y\sim f_Y(y\param\theta_0)$ for $\theta_0=(\phi_0,\psi_0)$, and $Z\sim N[0,1]$ is a normal random variable with mean $0$ and variance $1$.
In \eqref{LAN}, $\approx$ indicates approximate equality in distribution. 
Under regular asymptotics, the curvature of the quadratic approximation in LAN is the Fisher information, and LAN is therefore a similar property to asymptotic normality of the maximum likelihood estimate.
The quantity $I$ in \eqref{LAN} can be interpreted as the marginal Fisher information for $\phi$, also known as the $\phi$-aspect of the Fisher information \citep[][Section~3.4]{barndorff94}.
Specifically, if we write the inverse of the full Fisher information as
\begin{equation}\nonumber
V_\theta=\left[\begin{array}{cc}
V_{\phi} & V_{\phi\psi}\\
V_{\psi\phi} & V_\psi
\end{array}\right],
\end{equation}
then $I=V_\phi^{-1}$.
In this article, we focus on developing and demonstrating statistical methodology rather than on presenting theoretical results. 
Therefore, the formal mathematical representation of the approximations in this paper as asymptotic limit theorems is postponed to subsequent work.

The LAN property suggests that the Monte Carlo profile log likelihood evaluated at $\phi_{1:K}$ can be approximated, in a neighborhood of its maximum, by a quadratic metamodel, 
\begin{equation}\label{QuadraticMetamodel}
\breve \loglik^P_k(y) = -\hat a(y)\phi_k^2+\hat{b}(y)\phi_k+\hat{c}(y) + \epsilon_k, \quad\epsilon_k\sim\mathrm{IID}\big(\sigma^2(\data{y})\big), \quad k\in 1{\mycolon}K.
\end{equation}
This local quadratic metamodel is a special case of \eqref{ProfileMetamodelUnbiased}. 
The unknown coefficients $\hat a(\data{y})$, $\hat b(\data{y})$ and $\hat c(\data{y})$, corresponding to equation \eqref{QuadraticMetamodel} evaluated at $y=\data{y}$, describe a quadratic approximation to the numerically intractable likelihood surface.
We can use standard linear regression to estimate $\hat a(\data{y})$,  $\hat b(\data{y})$ and  $\hat c(\data{y})$ from the Monte Carlo profile evaluations.
Writing $\epsilon=\epsilon_{1:K}$, we denote the resulting linear regression coefficients as $\breve a^*=\breve a(\data{y},\epsilon)$, $\breve b^* = \breve b(\data{y},\epsilon)$ and $\breve c^* = \breve c(\data{y},\epsilon)$.
The Monte Carlo quadratic profile likelihood approximation is
\begin{equation}\label{qpla}
\breve \loglik^Q(\phi\param \data{y}) = -\breve a^* \phi^2 + \breve b^* \phi + \breve c^*.
\end{equation}
The marginal MLE $\hat\phi^*$ can be approximated by the maximum of $\breve \loglik^Q(\phi\param \data{y})$, which is given by
\begin{equation}  \nonumber
\breve\phi^Q(\data{y},\epsilon)=\frac{\breve b(\data{y},\epsilon)}{2\breve a(\data{y},\epsilon)}.
\end{equation}
Now, for $Y\sim f_Y(y\param\theta_0)$, we separate the variability in $\breve\phi^Q(Y,\epsilon)$ into two components:
\begin{enumerate}
\item {\it Statistical error} is the uncertainty resulting from randomness in the data, if the data were a draw from the statistical model. This is the error in the ideal quadratic profile approximation estimate $\hat b(\data{y})/2\hat a(\data{y})$ as an estimate of $\phi_0$. 

\item {\it Monte Carlo} error is the uncertainty resulting from implementing a Monte Carlo estimator. 
This is the error in $\breve b(\data{y},\epsilon)/2\breve a(\data{y},\epsilon)$ as a Monte Carlo estimate of $\hat b(\data{y})/2\hat a(\data{y})$.
\end{enumerate}
The LAN approximation in \eqref{LAN} suggests a normal approximation for the distribution of the marginal MLE $\hat\phi^*$ which we write as
\begin{equation} \label{ProfileCLT}
\hat\phi(Y)\approx N\big[\phi_0,\SE_{\,\mathrm{stat}}^2\big].
\end{equation}
The usual statistical standard error, $1\big/\sqrt{2\hat{a}(\data{y})}$, is not available to us, but we can instead use its Monte Carlo estimate,
\begin{equation} \label{SEstat}
\SE_{\,\mathrm{stat}} = \frac{1}{\sqrt{2\breve a(\data{y},\epsilon)}}.
\end{equation}
To quantify the Monte Carlo error, we first note that standard linear model methodology provides variance and covariance estimates $\breve\var\big[\breve a(\data{y},\epsilon)\big]$, $\breve\var\big[\breve b(\data{y},\epsilon)\big]$ and $\breve\cov\big[\breve a(\data{y},\epsilon),\breve b(\data{y},\epsilon)\big]$.
The regression errors representing only Monte Carlo variability conditional on $Y=\data{y}$, i.e., 
$\breve\var[\breve a(\data{y},\epsilon)]=\var[\breve a(Y,\epsilon)| Y=\data{y}]$.
A standard central limit approximation for regression coefficient estimates is
\begin{equation}  \nonumber
\left( \begin{array}{l} \breve a(\data{y},\epsilon) \\ \breve b(\data{y},\epsilon)\end{array}\right)
\approx
N\left[ 
\left(\begin{array}{l} \hat a(\data{y}) \\ \hat b(\data{y})\end{array}\right),
\left(\begin{array}{ll} \breve\var[\breve a(\data{y},\epsilon)] & \breve\cov[\breve a(\data{y},\epsilon),\breve b(\data{y},\epsilon)] \\
                  \breve\cov[\breve a(\data{y},\epsilon),\breve b(\data{y},\epsilon)] & \breve\var[\breve b(\data{y},\epsilon)] \end{array}\right) \right].
\end{equation}
An application of the delta method gives a central limit approximation for the maximum, conditional on $Y=\data{y}$, given by
\begin{equation} \label{DeltaMethod} 
\frac{\breve b(\data{y},\epsilon)}{2\breve a(\data{y},\epsilon)}
\approx
N\left[ 
\left(\frac{\hat b(\data{y})}{2\hat a(\data{y})}\right),
\SE_{\,\mathrm{mc}}^2
\right],
\end{equation}
where
\begin{eqnarray} \nonumber
&& \hspace{-3mm} \SE_{\,\mathrm{mc}}^2 = \breve\var\left[\frac{\breve b(\data{y},\epsilon)}{2\breve a(\data{y},\epsilon)}\right]
\\
\label{DeltaVariance}
&&\hspace{3mm}\approx  \frac{1}{4\breve a^2(\data{y},\epsilon)} \hspace{-1mm}
\left\{ 
\breve\var\big[\breve b(\data{y},\epsilon)]
- \frac{2\breve b(\data{y},\epsilon)}{\breve a(\data{y},\epsilon)}\breve\cov\big[\breve a(\data{y},\epsilon),\breve b(\data{y},\epsilon)\big]
+ \frac{\breve b^2(\data{y},\epsilon)}{\breve a^2(\data{y},\epsilon)}\breve\var\big[\breve a(\data{y},\epsilon)\big] \hspace{-1mm}
\right\}. \hspace{8mm}
\end{eqnarray}
To obtain the combined statistical and Monte Carlo error, we write 
\begin{equation} \label{ConditionalVariance}
\var\left[\frac{\breve b (Y,\epsilon)}{2\breve a(Y,\epsilon)}\right] = \E\left\{\var\left[\frac{\breve b (Y,\epsilon)}{2\breve a(Y,\epsilon)}\Big| Y\right]\right\}
+ \var\left\{\E\left[\frac{\breve b (Y,\epsilon)}{2\breve a(Y,\epsilon)} \Big| Y\right]\right\}.
\end{equation}
Now, from \eqref{LAN}, the curvature of the profile log likelihood is approximately constant, independent of $Y$. 
We suppose that the profile points used to obtain $\breve\phi^Q(Y,\epsilon)$ are approximately centered on $\breve\phi^Q(Y,\epsilon)$ regardless of the value of $Y$. 
This assumption can be satisfied by construction, for example by fitting the quadratic metamodel in \eqref{QuadraticMetamodel} using local weights (as in the MCAP algorithm below).
Further, we suppose that $\var[\epsilon_k(Y)]\approx\var[\epsilon_k(\data{y})]$.
Together, these approximations imply
\begin{equation}  \label{EV}
\var\left[\frac{\breve b (Y,\epsilon)}{2\breve a(Y,\epsilon)}\Big| Y\right]
\approx
\var\left[\frac{\breve b(\data{y},\epsilon)}{2\breve a(\data{y},\epsilon)}\right].
\end{equation}
Also, from the central limit approximation in \eqref{DeltaMethod}, we have
\begin{equation}  \label{VE}
\E\left[\frac{\breve b (Y,\epsilon)}{2\breve a(Y,\epsilon)} \Big| Y\right] \approx \hat\phi(Y).
\end{equation}
Putting \eqref{EV} and \eqref{VE} into \eqref{ConditionalVariance}, and using the approximations in \eqref{ProfileCLT} and \eqref{DeltaMethod}, we get
\begin{equation} \nonumber
 \frac{\breve b(Y,\epsilon)}{2\breve a(Y,\epsilon)}
\approx
N\left[ \phi, \SE^2_{\,\mathrm{total}}\right],
\end{equation}
where
\begin{equation}  \nonumber
\SE^2_{\,\mathrm{total}} = \sqrt{\SE_{\,\mathrm{mc}}^2 + \SE_{\,\mathrm{stat}}^2}.
\end{equation}
The usual asymptotic profile likelihood confidence interval cutoff value can be obtained by converting the standard error of the MLE into an equivalent cutoff on a quadratic approximation to the profile log likelihood.
In our setting, the asymptotic $(1-\alpha)$ confidence interval, $\breve\phi^Q \pm z_\alpha \times \SE_{\,\mathrm{total}}$ where $\prob[Z>z_\alpha]=\alpha/2$, is equivalent to a Monte Carlo adjusted profile cutoff for the quadratic approximation $\breve\loglik^Q(\phi\param\data{y})$ of
\begin{equation} \label{Delta}
\delta = \breve a^* \times \big( z_\alpha \times {\SE}_{\,\mathrm{total}}\big)^2 = z_\alpha^2\left(\breve a^* \times {\SE}_{\,\mathrm{mc}}^2+ \frac{1}{2}\right).
\end{equation}
Note that, if ${\SE}_{\,\mathrm{mc}}=0$, the calculation in \eqref{Delta} for $\alpha=0.05$ reduces to
\begin{equation}\nonumber
\delta = 1.96^2/2 = 1.92,
\end{equation} 
the usual cutoff to construct a 95\% confidence interval for an exact profile likelihood.

Confidence intervals based on a quadratic approximation to the exact log likelihood are asymptotically equivalent to using the same cutoff $\delta$ with a smoothed version of the likelihood, so long as an appropriate smoother is used \citep{ionides05}.
An appropriate smoother should return a quadratic when the points do indeed lie on a quadratic, a property satisfied, for example, by local quadratic smoothing such as the R function \texttt{loess}.
We therefore propose using $\delta$ as an appropriate cutoff on a profile likelihood estimate obtained by applying a suitable smoother to the Monte Carlo evaluations in \eqref{ProfileMetamodel}.
A smoother, $S\big(\phi\param \phi_{1:K},\breve \loglik^P_{1:K},\lambda\big)$, generates a value at $\phi$ based on fitting a smooth curve through the points $\big\{(\phi_k,\breve\loglik^P_{k}),k\in 1{\mycolon}K\big\}$ with an algorithmic parameter $\lambda$ determining the smoothness of the fit.
A resulting maximum smoothed Monte Carlo profile likelihood estimate is
\begin{equation}
\breve\phi^S=\arg\max_\phi S\big(\phi\param \phi_{1:K},\breve\loglik^P_{1:K},\lambda\big).
\end{equation}
A corresponding Monte Carlo profile likelihood confidence interval for a cutoff $\delta$ is 
\begin{equation}
\big\{\phi: S\big(\phi\param \phi_{1:K},\breve\loglik^P_{1:K},\lambda\big) >
 S\big(\breve\phi^S\param \phi_{1:K},\breve\loglik^P_{1:K},\lambda\big)-\delta\big\}.
\end{equation}
Here, we suppose that $S\big(\phi\param \phi_{1:K},\breve\loglik^P_{1:K},\lambda\big)$ is evaluated at $\phi$ via local quadratic regression with weight $w_{k}(\phi)$ on the point $\big(\phi_k,\breve\loglik^P_k\big)$, where $w_{k}(\phi)$ depends on the proximity of $\phi$ to $\phi_k$.
Specifically, we take $S$ to be the widely used local quadratic smoother of \citet{cleveland93} as implemented in the R function \texttt{loess}.
In this case, $\lambda$ is the {\it span} of the smoother, defined as the fraction of the data used to construct the weights in the local regression at any point $\phi$.
In practice, the statistician needs to specify $\lambda$. 
While automated choices of smoothing parameter have been proposed, it remains standard practice to choose the smoothing parameter based on some experimentation and looking at the resulting fit.
In our experience, the default \texttt{loess} choice of $\lambda=0.75$ has been appropriate in most cases.
However, a larger value of $\lambda$ is needed when the profile is evaluated at very few points (as demonstrated in Section~\ref{sec:spacetime}).
When the exact profile is not far from quadratic, one can expect local quadratic smoothing of the Monte Carlo profile likelihood to be insensitive to the choice of $\lambda$.

Just as the local quadratic regression smoother has weights $w_{1:K}(\phi)$, the quadratic metamodel in \eqref{QuadraticMetamodel} can be fitted using regression weights.
A natural choice of these weights for obtaining a profile confidence interval cutoff for $S\big(\phi\param \phi_{1:K},\breve\loglik^P_{1:K},\lambda\big)$ is $w_{1:K}(\breve\phi^S)$.
This choice is used for the MCAP algorithm below.
For our numerical results, we used the implementation of this MCAP algorithm given in a supplement (Section~\ref{sec:mcap:code}).

\vspace{2mm}

\newcommand\mystretch{\rule[-2mm]{0mm}{5mm} }   
\newcommand\asp{\hspace{4mm}}

\begin{center}
\noindent\begin{tabular}{l}
\hline
{\bf Algorithm~{MCAP} (Monte Carlo adjusted profile)
}\rule[-1.5mm]{0mm}{6mm}\\
\hline
{\bf input:}\rule[-1.5mm]{0mm}{6mm} \\
Monte Carlo profile $\breve\loglik^P_{1:K}$ evaluated at $\phi_{1:K}$\\
Local quadratic regression smoother, $S$\\
Smoothing parameter, $\lambda$\\
confidence level, $1-\alpha$\\
{\bf output:}\rule[-1.5mm]{0mm}{6mm} \\
Cutoff, $\delta$, for a Monte Carlo profile likelihood confidence interval\\
\hline
Fit a local quadratic smoother, $\breve\loglik^{S}(\phi)=S(\phi\param \phi_{1:K}, \breve\loglik^P_{1:K},\lambda)$\\
Obtain $\breve\phi^S=\arg\max \breve\loglik^{S}(\phi)$\\
Obtain regression weights $w_{1:K}$ for the evaluation of $S(\phi\param \phi_{1:K}, \breve\loglik^P_{1:K},\lambda)$ at $\phi=\breve\phi^S$\\
Fit a linear regression model, $\breve\loglik^P_k = -a \phi_k^2 + b\phi_k + c + \epsilon_k$, with weights $w_{1:K}$\\
Obtain regression estimates $\breve a$ and $\breve b$\\
Obtain regression covariances $\breve\var[\breve a]$, $\breve\var[\breve b]$, $\breve\cov[\breve a,\breve b]$\\
Let $\SE_{\,\mathrm{mc}}^2 =\frac{1}{4\breve a^2}
\left\{ 
\breve\var[\breve b]
- \frac{2\breve b}{\breve a}\breve\cov[\breve a,\breve b]
+ \frac{\breve b^2}{\breve a^2}\breve\var[\breve a] 
\right\}$\\
Let $\chi_\alpha$ be the $(1-\alpha)$ quantile of the chi-square distribution on one degree of freedom\\
Let $\delta = \chi_\alpha\left(\breve a \times {\SE}_{\,\mathrm{mc}}^2+ 1/2\right)$ \\
\hline
\end{tabular}
\end{center}


\section{Example: inference for partially observed dynamic systems} 
\label{sec:pomp}

Many dynamic systems with indirectly observed latent processes can be modeled within the partially observed Markov process (POMP) framework.
A general POMP model, also known as a hidden Markov model or a state space model, consists of a latent Markov process $\{X(t)\}$, with $X(t)$ taking values in a space $\Xspace$, together with a sequence of observable random variables $Y_1,\dots,Y_N$.
We suppose $Y_n$ occurs at a time $t_n$, and the observations are conditionally independent 
of each other and of $\{X(t)\}$ given $X(t_1),\dots,X(t_N)$.
For example, we may have $\Yspace=\R^d$, the space of $d$-dimensional real vectors.  
When $d=1$ (or $d$ is small) $Y_1,\dots,Y_N$ is called a univariate (or multivariate) time series model.
The POMP framework provides a fundamental approach for nonlinear time series analysis, with innumerable applications \citep{breto09}.
When $d$ becomes large, the POMP framework allows for nonlinear panel data and spatiotemporal data, as well as other complex data structures.
Unless the POMP model is linear and Gaussian, or $\Xspace$ is a sufficiently small finite set, Monte Carlo techniques such as sequential Monte Carlo (SMC) are required to evaluate the likelihood function.
For our examples, we focus on likelihood maximization by iterated filtering \citep{ionides15}.
Similar issues arise with alternative computational approaches, including Monte~Carlo Expectation-Maximization algorithms \citep[][Chapter~11]{cappe05}.
Even for the relatively simple case of time series POMP models (discussed further in Section~\ref{sec:timeseries}) numerical issues can be computationally demanding for currently available methodology, giving opportunity for MCAP methodology to facilitate data analysis.
However, to demonstrate the capabilities of our methodology, we present three high-dimensional POMP inference challenges that become computationally tractable using MCAP.

\subsection{Inferring population dynamics from genetic sequence data}\label{sec:genPomp}

Genetic sequence data on a sample of individuals in an ecological system has potential to reveal population dynamics.
Extraction of this information has been termed {\it phylodynamics} \citep{grenfell04}.
Likelihood-based inference for joint models of the molecular evolution process, population dynamics, and measurement process is a challenging computational problem.
The bulk of extant phylodynamic methodology has therefore focused on inference for population dynamics conditional on an estimated phylogeny and replacing the population dynamic model with an approximation, called a {\it coalescent model} that is convenient for calculations backwards in time \citep{karcher16}.
Working with the full joint likelihood is not entirely beyond modern computational capabilities; in particular it can be done using the {\genPomp} algorithm of \citet{smith16}.
The {\genPomp} algorithm is an application of iterated filtering methodology \citep{ionides15} to phylodynamic models and data.
To the best of our knowledge, {\genPomp} is the first algorithm capable of carrying out full joint likelihood-based inference for population-level phylodynamic inference.
However, the {\genPomp} algorithm leads to estimators with high Monte Carlo variance, indeed, too high for reasonable amounts of computation resources to reduce Monte Carlo variability to negligibility.
This, therefore, provides a useful scenario to demonstrate our methodology.

\begin{figure}
\begin{knitrout}\small
\definecolor{shadecolor}{rgb}{0.969, 0.969, 0.969}\color{fgcolor}

{\centering \includegraphics[width=\maxwidth]{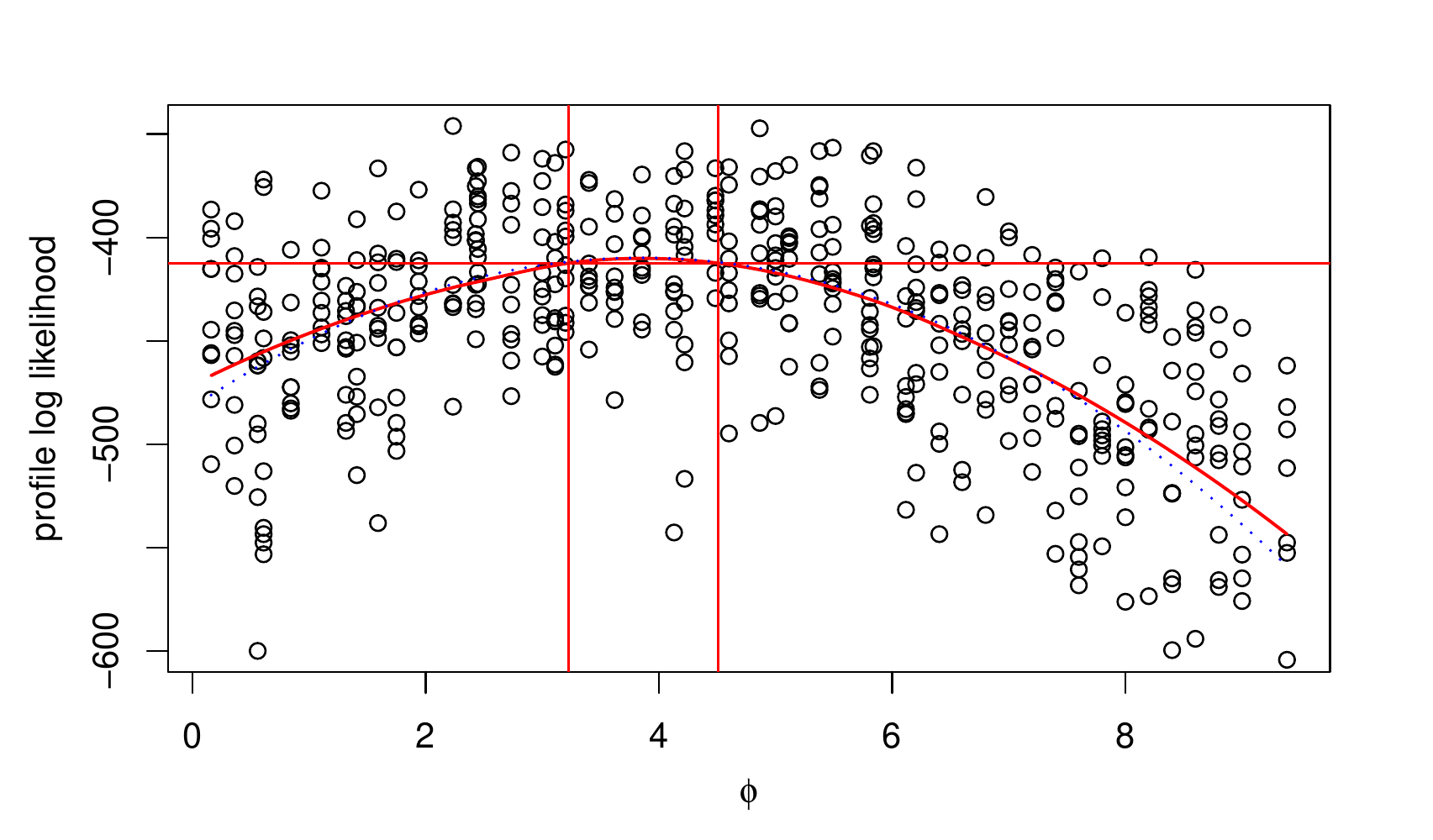} 

}

\end{knitrout}
\caption{Profile likelihood for an infectious disease transmission parameter inferred from genetic data on pathogens. 
The smoothed profile likelihood and corresponding MCAP 95\% confidence interval are shown as solid red lines. 
The quadratic approximation in a neighborhood of the maximum is shown as a blue dotted line.
}\label{fig:gen_profile}
\end{figure}

Figure~\ref{fig:gen_profile} presents a Monte Carlo profile computed by \citet{smith16}, with confidence intervals constructed by applying the MCAP algorithm implemented by the \texttt{mcap} procedure (Section~\ref{sec:mcap:code}) with default smoothing parameter.
The model and data concern HIV transmission in Southeast Michigan, but details of the model and computations are not of immediate interest since all we need to consider are the estimated profile likelihood points.
The profiled parameter quantifies HIV transmission from recently infected, diagnosed individuals---it is $\varepsilon_{J_0}$ in the notation of \citet{smith16} but we rename it as $\phi$ for the current paper.
The computations for Figure~\ref{fig:gen_profile} took approximately 10 days using 500 cores on a Linux cluster.
To scale this methodology to increasingly large datasets and more complex models, it is apparent that one may be limited by the computational effort required to control Monte Carlo error.
The MCAP procedure gives a Monte Carlo standard error of $\mathrm{SE}_{\,\mathrm{mc}}=0.151$ on the value maximizing the smoothed Monte Carlo profile, based on the quadratic approximation at the maximum.
The statistical error is $\mathrm{SE}_{\,\mathrm{stat}}=0.32$. 
Combining these sources of uncertainty gives a total standard error of $\mathrm{SE}_{\,\mathrm{total}}=0.354$.
From \eqref{Delta}, the resulting $95\%$ confidence profile cutoff is $\delta = 2.35$.
We see in Figure~\ref{fig:gen_profile} that the smoothed profile is close to its quadratic approximation in the neighborhood of the maximum statistically supported by the data.
We also see that the Monte Carlo uncertainty in the profile confidence interval is rather small, leading to a profile cutoff not much bigger than the value of 1.92 for zero Monte Carlo error, despite the large Monte Carlo variability in the evaluation of any one point on the profile.

\subsection{Panel time series analysis}\label{sec:panelPomp}

Panel data consists of a collection of time series which have some shared parameters, but negligible dynamic dependence.
We consider inference using mechanistic models for panel data, i.e., equations for how the process progresses through time derived from scientific principles about the system under investigation.
In principle, statistical methods for mechanistic time series analysis \citep{breto09} extend to the panel situation \citep{breto16}.
However, extensive data add computational challenges to Monte Carlo inference schemes.
In particular, with increasing amounts of data, it must eventually become infeasible to calculate the likelihood with an error as small as one log unit.
The MCAP procedure nevertheless succeeds so long as the signal-to-noise ratio in the Monte Carlo profile is adequate.
In a simple situation, where each time series is modeled as independent and identically distributed and each time series model contains the same parameters, we can check how this ratio scales.
The Fisher information scales linearly with the number of time series in the panel, and therefore the curvature of the log likelihood profile also scales linearly.
The Monte Carlo standard error on the likelihood scales at a square-root rate.
In this case, we therefore expect the MCAP methodology to scale successfully with the number of time series in the panel.

Investigations of population-level infectious disease transmission lead to highly nonlinear, stochastic, partially observed dynamic models.
The great majority of disease transmission is local, despite the importance of spatial transmission to seed the local epidemics \citep{bjornstad07}.
Fitting models to panels of epidemiological time series data, such as incidence data for collections of cities or states, offers potential to elucidate the similarities and differences between these local epidemics.  

\begin{figure}
\begin{knitrout}\small
\definecolor{shadecolor}{rgb}{0.969, 0.969, 0.969}\color{fgcolor}

{\centering \includegraphics[width=\maxwidth]{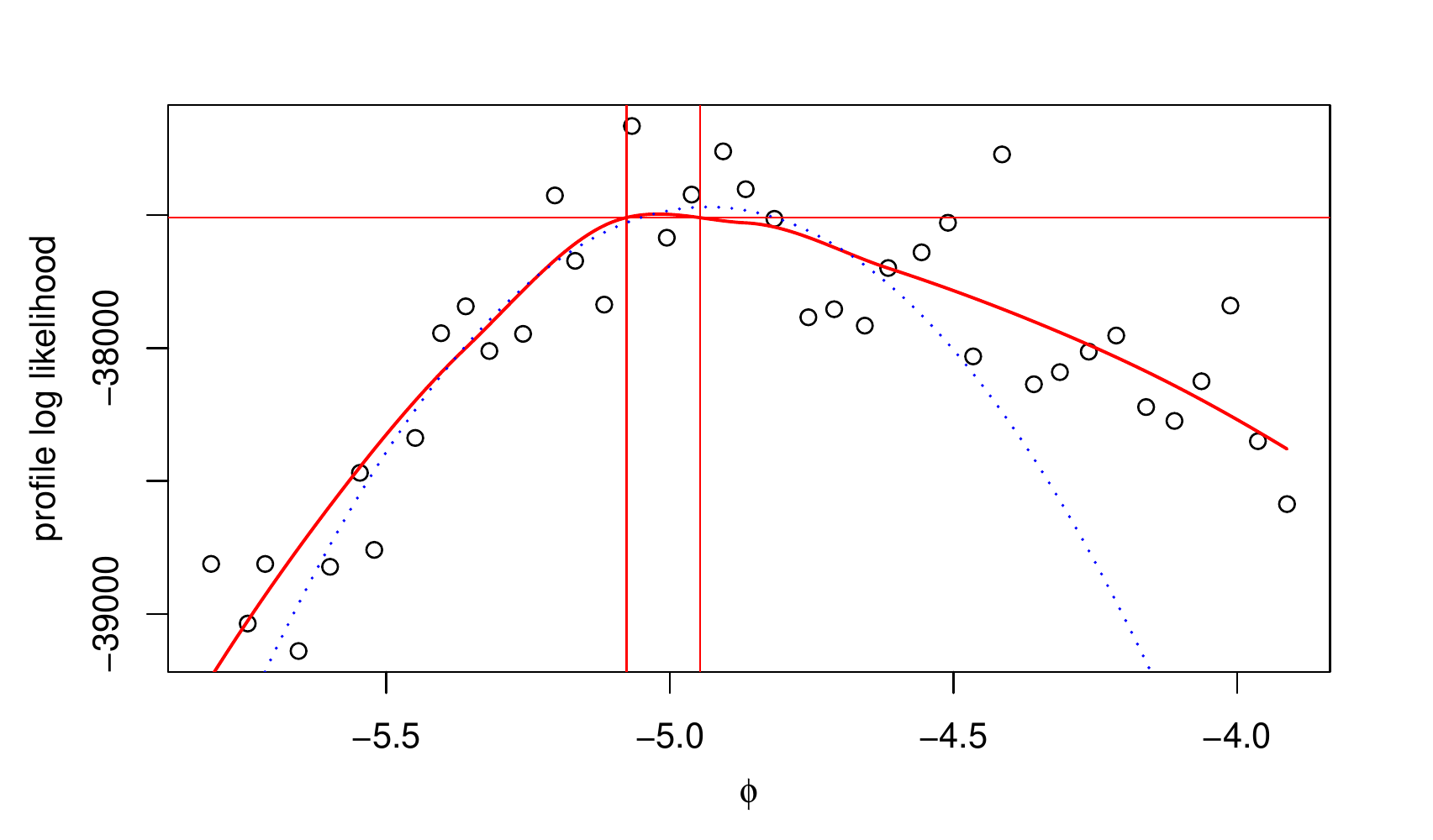} 

}

\end{knitrout}
\caption{Profile likelihood for a nonlinear partially observed Markov process model for a panel of time series of historical state-level polio incidence in the United States. 
The smoothed profile likelihood and corresponding MCAP 95\% confidence interval are shown as solid red lines. 
The quadratic approximation in a neighborhood of the maximum is shown as a dotted blue line.}\label{fig:panel_profile}
\end{figure}

We demonstrate the MCAP procedure on a panel estimate of the reporting rate of paralytic polio in the pre-vaccination era United States. 
Reporting rate has important consequences for understanding the system: conditional on observed incidence data, reporting rate determines the extent of the unreported epidemic.
Yet, in the presence of many uncertainties about this complex disease transmission system, a single disease incidence time series often cannot conclusively pin down this epidemiological parameter.
The profile evaluations in Figure~\ref{fig:panel_profile} were obtained by \citet{breto16} in an extension of the analysis of \citet{martinez-bakker15}.
\citet{martinez-bakker15} analyzed state level paralytic polio incidence data in order to study the role of unobserved asymptomatic polio infections in disease persistence. 
Here, the reporting rate parameter \citep[$\log(\rho)$ in the terminology of][]{breto16} is denoted by $\phi$.
The MCAP procedure gives a Monte Carlo standard error of $\mathrm{SE}_{\,\mathrm{mc}}=0.033$ and a statistical error of $\mathrm{SE}_{\,\mathrm{stat}}=0.013$. Combining them gives a total standard error of $\mathrm{SE}_{\,\mathrm{total}}=0.035$. 
The resulting profile cutoff is $\delta = 13.6.$
The profile decreases slowly to the right of the MLE, since higher reporting rates can be compensated for by lower transmission intensities.
The model struggles to explain reporting rates much lower than the MLE, since the reporting rate must be sufficient to explain the observed number of cases in a situation where almost all individuals acquire non-paralytic polio infections.
This asymmmetrical tradeoff may explain why the profile log likelihood shows some noticeable deviation from its quadratic approximation in a neighborhood of the maximum.

The computations for Figure~\ref{fig:panel_profile} required approximately 24 hours on 300 cores. 
At this level of computational intensity, we see that the majority of uncertainty about the parameter $\phi$ is due to Monte Carlo error rather than statistical error. 
For this large panel dataset, in the context of the fitted model, the parameter $\phi$ would be identified very accurately by the data if we had access to the actual likelihood surface.
Additional computation could, therefore, reduce the uncertainty on our estimate of $\phi$ by a factor of three.
However, the data analyst may decide the available computational effort is better used exploring other parameters or alternative model specifications.

\subsection{Inference for nonlinear partially observed spatiotemporal systems}\label{sec:spacetime}

\begin{figure}
\begin{knitrout}\small
\definecolor{shadecolor}{rgb}{0.969, 0.969, 0.969}\color{fgcolor}

{\centering \includegraphics[width=\maxwidth]{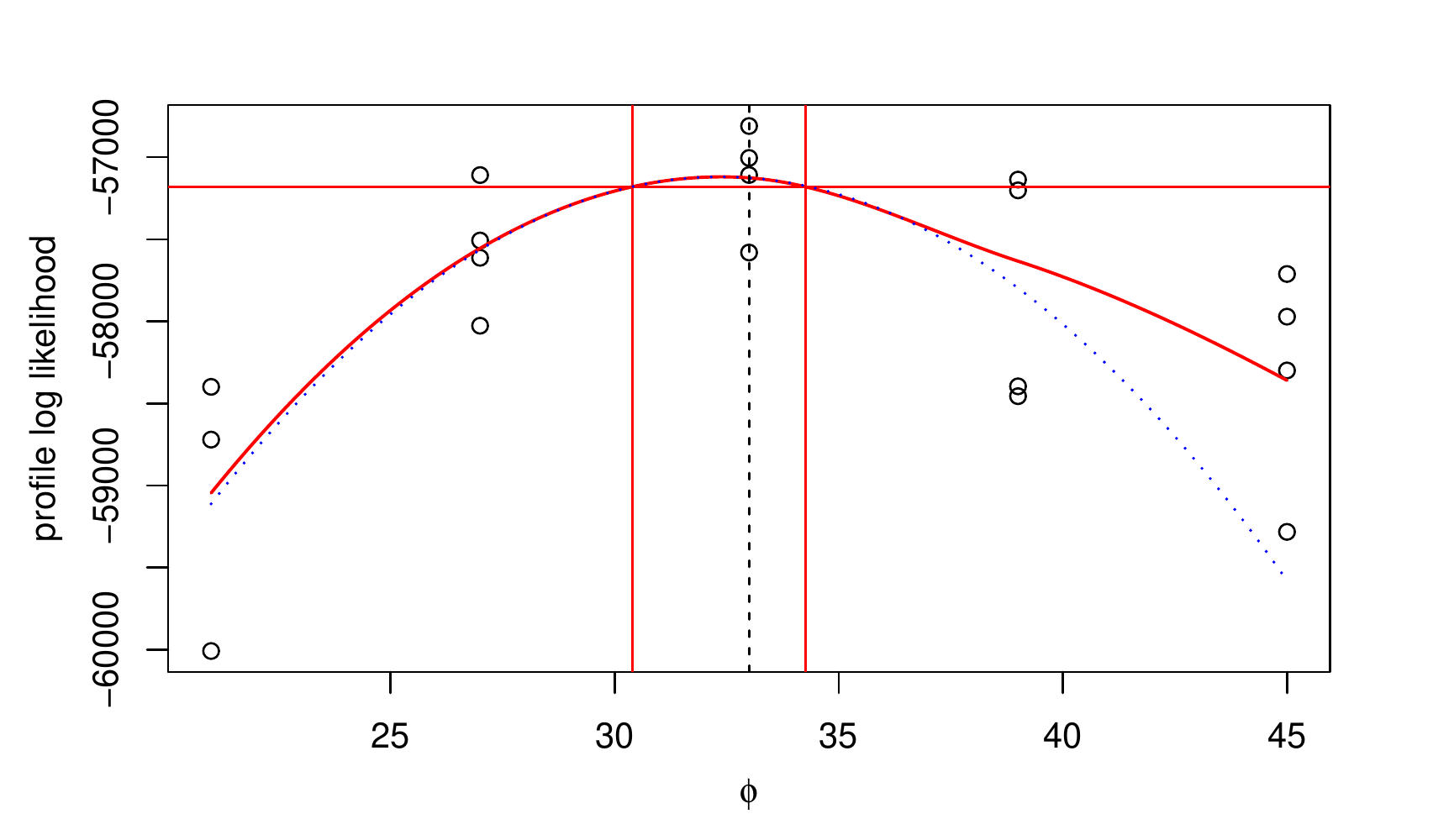} 

}

\end{knitrout}
\caption{Profile likelihood for a spatiotemporal measles transmission model with twenty metapopulations.
The profile parameter describes the contact rate within each metapopulation.
For this figure, the data were simulated from a fitted model, and so the true parameter can be shown (dashed black vertical line).
The smoothed profile likelihood and corresponding MCAP  95\% confidence interval are shown as solid red lines. 
The quadratic approximation in a neighborhood of the maximum is shown as a dotted blue line.
}\label{fig:stif}
\end{figure}

Spatiotemporal data consists of time series collected at various locations.
Spatiotemporal models extend panel models by allowing for dynamic dependence between locations.
We consider statistical inference for a mechanistic spatiotemporal model, meaning a collection of nonlinear partially observed spatially coupled Markov process.
SMC methods, that provide a foundation for much likelihood-based inference relating POMP models to time series data, struggle with spatiotemporal data since they scale poorly with spatial dimension \citep{bengtsson08}.
Theoretically, SMC methods with sub-exponential scaling can be developed for weakly coupled spatiotemporal systems \citep{rebeschini15}.
In practice, appropriately designed SMC schemes can successfully carry out Monte Carlo likelihood evaluation for general partially observed spatiotemporal processes of modest dimension \citep{park16}.
\citet{park16} then employed iterated filtering methodology \citep{ionides15} which modifies an SMC algorithm to maximize the likelihood.
Figure~\ref{fig:stif} shows an estimated likelihood profile for a parameter $\phi$ corresponding to the contact rate between individuals  \citep[denoted as $\beta$ by][]{park16} when fitting a ten parameter model to pre-vaccination measles incidence in 20 cities in the United Kingdom. 
This profile corresponds to a simulation test of the methodology of \citet{park16} in which the true parameter is known.
Here, we are not immediately concerned with the details of the model and the Monte Carlo methodology \citep[described by][]{park16} but rather with showing another example of how a computationally demanding inference problem can give rise to noisy Monte Carlo points estimating a profile likelihood.
For this computation, only five distinct parameter values were used when computing the profile. 
The default smoothing parameter $\lambda=0.75$ was too small in this case, since the local quadratic fit by the smoother at the maximum placed almost all its weights on only three distinct parameter values.
The resulting numerical instability was avoided by taking $\lambda=1$.
For this analysis, the profile cutoff adjusted for Monte Carlo uncertainty is $\delta=61.6$, and we see that the Monte Carlo variability $\mathrm{SE}_{\,\mathrm{mc}}=1.00$ in the parameter estimate greatly exceeds the statistical variability $\mathrm{SE}_{\,\mathrm{stat}}=0.18$.
Evidently, the simulated spatiotemporal data have a considerable amount of information about the parameter $\phi$, but extracting that information in a statistically efficient way is complicated by the computational challenge of working with the likelihood of a nonlinear partially observed spatiotemporal process. 

\subsection{Applications to time series analysis via mechanistic models}
\label{sec:timeseries}

The examples in Sections~\ref{sec:genPomp}, \ref{sec:panelPomp} and~\ref{sec:spacetime} demonstrate applications which were computationally intractable without MCAP.
Applications of the POMP framework to nonlinear time series analysis typically involve smaller data sets, and a relatively simple dependence structure, and are therefore less computationally demanding.
This consideration has facilitated the utilization of Monte Carlo profile likelihood, without the benefits of MCAP, as a technique at the cutting edge of nonlinear time series analysis.
In the context of infectious disease dynamics, \citet{dobson14} wrote,
``Powerful new inferential fitting methods  \citep{ionides06-pnas} considerably increase the accuracy of outbreak predictions while also allowing models whose structure reflects different underlying assumptions to be compared. 
These approaches move well beyond time series and statistical regression analyses as they include mechanistic details as mathematical functions that define rates of loss of immunity and the response of vector abundance to climate.''
Examples showing a  central role for Monte Carlo profile likelihood in such analyses are given by \citet[][Fig.~2]{king08}, \citet[][Figs.~S3 and S8A]{camacho11}, \citet[][Fig.~3A]{blackwood13}, \citet[Figs.~2B-2G and 4L-4P]{shrestha13} and \citet[Figs.~S1, S4 and S5]{blake14}.
The main practical limitation of this approach is computational resources \citep{he10}.
We have shown that our methodology can both quantify and dramatically reduce the Monte Carlo error in computationally intensive inferences for POMP models.
The MCAP procedure therefore improves the accessibility and scalability of inference for nonlinear time series models.

\section{A simulation study of the MCAP procedure}\label{sec:sim}

We look for a numerically convenient toy scenario that generates Monte Carlo profiles resembling Figures~\ref{fig:gen_profile}, ~\ref{fig:panel_profile} and~\ref{fig:stif}.
Our simulated data are an independent, identically distributed log normal sample $Y_{1:N}$ where $\log(Y_n)\sim N[\phi,2\sigma^2]$ for $n\in 1{\mycolon}N$. 
We consider a profile likelihood confidence interval for the log mean parameter, $\phi$.
The log normal distribution leads to log likelihood profiles that deviate from quadratic.
To set up a situation with Monte Carlo error in evaluating and maximizing the likelihood, we supposed that the likelihood is accessed via Monte Carlo integration of a latent variable.
Specifically, we write $Y_n|X_n\sim \mathrm{lognormal}(X_n, \sigma^2)$ with $X_n\sim N[\phi,\sigma^2]$.
Then, our Monte Carlo density estimator is
\begin{equation}\label{eq:mc-density}
\breve f_Y(y\param \phi,\sigma,s,J)=\frac{1}{J}\sum_{j=1}^J f_{LN}(y\param \phi+\sigma\epsilon_j,\sigma^2),
\end{equation}
where $f_{LN}(y\param\mu,\tau^2)$ is the log normal density,
\begin{equation}\nonumber
f_{LN}(y\param\mu,\tau^2)=\frac{1}{y\tau\sqrt{2\pi}}
\exp\left\{\frac{-(\log y - \mu)^2}{2\tau^2}\right\},
\end{equation}
and $\epsilon_{1:J}$ is a sequence of standard normal pseudo-random numbers corresponding to a seed $s$.
We suppose that we are working with a parallel random number generator such that pseudo-random sequences corresponding to different seeds behave numerically like independent random sequences.
Our Monte Carlo log likelihood estimator is
\begin{equation}\label{eq:mc-loglik}
\breve \loglik(\phi,\sigma\param y_{1:N},s,J)= \sum_{n=1}^N \log \breve f_Y(y_n\param\phi,\sigma,s+n-1,J).
\end{equation}
Our Monte Carlo profile is calculated at $\phi\in\phi_{1:K}$. 
We maximize the likelihood numerically, at a fixed seed, to give a corresponding estimate of $\sigma$ given by 
\begin{equation}\label{eq:toy:sigma}
\breve \sigma^P_k(y_{1:N},s,J)=\arg\max_\sigma  \loglik\big(\phi_k,\sigma\param y_{1:N},s+N(k-1),J\big).
\end{equation}
We do not wish to imply that practical examples will generally result from a fixed-seed Monte Carlo likelihood calculation. 
Seed fixing is an effective technique for removing Monte Carlo variability from relatively small calculations, but can become difficult or impossible to implement effectively for complex, coupled, nonlinear systems.

\begin{figure}
\begin{knitrout}\small
\definecolor{shadecolor}{rgb}{0.969, 0.969, 0.969}\color{fgcolor}

{\centering \includegraphics[width=\maxwidth]{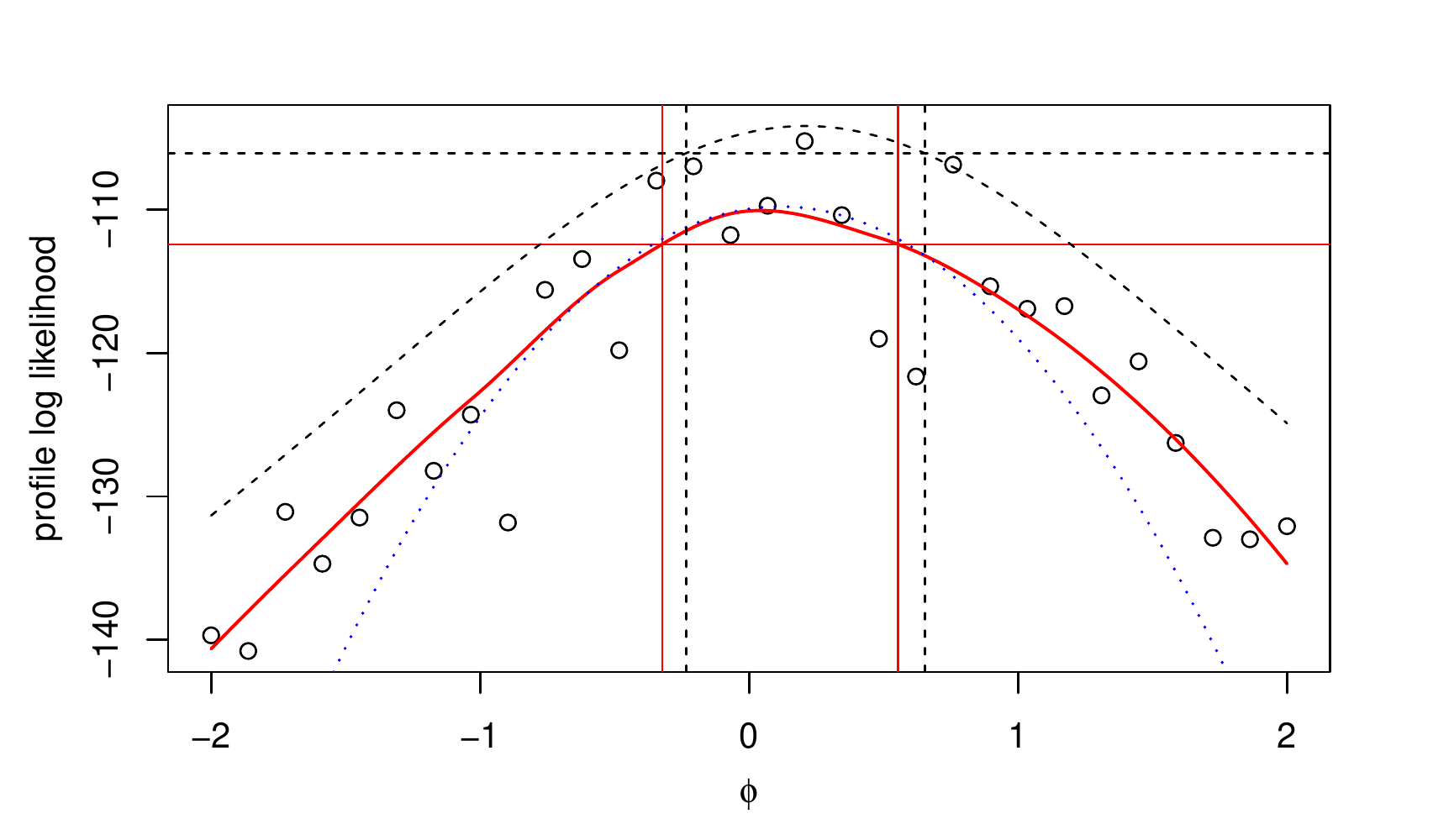} 

}

\end{knitrout}
\caption{Profile construction for the toy model. The exact profile and its asymptotic 95\% confidence interval are constructed with black dashed lines. 
Points show Monte Carlo profile evaluations. 
The MCAP is constructed in solid red lines, using the default $\lambda=0.75$ smoothing parameter.
The quadratic approximation used to calculate the MCAP profile cutoff is shown as a dotted blue line.}\label{fig:toy-plot}
\end{figure}

The following numerical results used $N=50$ and $J=3$ with true parameter values $\phi_0=0$ and $\sigma^2_0=1$.
There are two ways to increase the Monte Carlo error in the log likelihood for this toy example, by increasing the sample size, $N$, and decreasing the Monte Carlo effort, $J$.
The Monte Carlo variance of the log likelihood estimate increases linearly with $N$, but at the same time the curvature of the log likelihood increases and, within the inferentially relevant region, the profile log likelihood becomes increasingly close to quadratic.
Thus, in the context of our methodology, increasing $N$ actually makes inference easier despite the increasing Monte Carlo noise.
This avoids a paradoxical difficulty of Monte Carlo inference for big data: more data should be a help for a statistician, not a hindrance!
Decreasing $J$ represents a situation where Monte Carlo variability increases without increasing information about the parameter of interest. 
In this case, the Monte Carlo variability and the Monte Carlo bias on the log likelihood due to Jensen's inequality both increase.
Also, likelihood maximization becomes more erratic for small $J$ since the maximization error due to the fixed seed becomes more important. 
However, Figure~\ref{fig:toy-plot} shows that, even when there is considerable bias and variance in the Monte Carlo profile evaluations, the Monte Carlo profile confidence intervals can be little wider than the exact interval. 

We computed intervals with nominal coverage of 95\%.
The MCAP coverage here was $93.4$\%, compared to $94.3$\% for the asymptotically exact profile (with a simulation study Monte Carlo standard error of $0.2$\%).
The MCAP intervals were, on average, $12.5$\% larger than the corresponding exact profile interval, with the increased width accounting for the additional Monte Carlo uncertainty.

\section{Discussion}\label{sec:discussion}

This paper has focused on likelihood-based confidence intervals.
An alternative to likelihood-based inference is to compare the data with simulations using some summary statistic. 
Various plug-and-play methodologies of this kind have been proposed, such as synthetic likelihood \citep{wood10} and nonlinear forecasting \citep{ellner98}.
For large nonlinear systems, it can be hard to find low-dimensional summary statistics that capture a good fraction of the information in the data. 
Even summary statistics derived by careful scientific or statistical reasoning have been found surprisingly uninformative compared to the whole data likelihood in both scientific investigations \citep{shrestha11} and simulation experiments \citep{fasiolo16}.

Much attention has been given to scaling Bayesian computation to complex models and large data.
Bayesian computation is closely related to likelihood inference for stochastic dynamic models: the random variables generating a dynamic system are typically not directly observed, and these latent random variables are therefore similar to Bayesian parameters. 
We refer to these latent random variables as {\em random effects} since they have a similar role as linear model random effects.
To carry out inference on the structural parameters of the model (i.e., the vector $\theta$ in this article) the Bayesian approach looks for the marginal posterior of $\theta$, which involves integration over the random effects.
Likelihood-based inference for $\theta$ similarly involves integrating out the random effects.
Numerical methods such as expectation propagation (EP) \citep{gelman14} and variational Bayes \citep{hoffman13} are effective for some model classes.
Another approach is to combine Markov chain Monte Carlo (MCMC) computations on subsets of the data, as in the posterior interval estimation (PIE) method of \citet{li16}.
The above approaches (EP, VB and PIE) all emphasize situations where the joint density of the data and latent variables can be conveniently split up into conditionally independent chunks, such as a hierarchical model structure.
Our methodology has no such requirement.
The panel model example above does have a natural hierarchical structure, with individual panels being independent (in the frequentist model sense) or conditionally independent given the shared parameters (in the Bayesian model sense).
Our spatiotemporal and genetic examples do not have such a representation.

Some simulation-based Bayesian computation methodologies have built on the observation that unbiased Monte Carlo likelihood computations can be used inside an MCMC algorithm \citep{andrieu09}.
For large systems, high Monte Carlo variability of likelihood estimates is a concern, in this context, since it slows down MCMC convergence \citep{bardenet15}.
\citet{doucet15} found that, for a given computational budget, the optimal balance between number of MCMC iterations and time spent on each likelihood evaluation occurs at a Monte Carlo likelihood standard deviation of one log unit.
For the systems we demonstrate, Monte Carlo errors that small are not computationally feasible.

Our simple and general approach permits inference when the signal-to-noise ratio in the Monte Carlo profile log likelihood is sufficient to uncover the main features of this function, up to an unimportant vertical shift.
For large datasets in which the signal (quantified as the curvature of the log likelihood) is large, the methodology can be effective even when the Monte Carlo noise is far too big to carry out standard MCMC techniques. 
Although the frequentist motivation for likelihood-based inference differs from the goal of Bayesian posterior inference, both approaches can be used for deductive scientific reasoning \citep{gelman13,ionides16}.

\vspace{4mm}

\ifblinded 
{ }
\else
{
\noindent{\Large\bf Acknowledgements}

\vspace{3mm}

\noindent This research was supported by National Science Foundation grant DMS-1308919 and National Institutes of Health grants 1-U54-GM111274 and 1-U01-GM110712.
}
\fi


\clearpage

\renewcommand{\contentsname}{Supplementary Content}
\renewcommand{\refname}{Supplementary References}
\renewcommand\thefigure{S-\arabic{figure}}
\renewcommand\thetable{S-\arabic{table}}
\renewcommand\thepage{S-\arabic{page}}
\renewcommand\thesection{S\arabic{section}}
\renewcommand\theequation{S\arabic{equation}}

\setcounter{section}{0}
\setcounter{page}{1}
 
\section{Supplement: Implementation of the MCAP algorithm in R}\label{sec:mcap:code}

The following R code carries out the MCAP algorithm, as used for the results in this paper.

\begin{knitrout}\small
\definecolor{shadecolor}{rgb}{0.969, 0.969, 0.969}\color{fgcolor}\begin{kframe}
\begin{alltt}
\hlstd{mcap} \hlkwb{<-} \hlkwa{function}\hlstd{(}\hlkwc{lp}\hlstd{,}\hlkwc{parameter}\hlstd{,}\hlkwc{confidence}\hlstd{=}\hlnum{0.95}\hlstd{,}\hlkwc{lambda}\hlstd{=}\hlnum{0.75}\hlstd{,}\hlkwc{Ngrid}\hlstd{=}\hlnum{1000}\hlstd{)\{}
  \hlstd{smooth_fit} \hlkwb{<-} \hlkwd{loess}\hlstd{(lp} \hlopt{~} \hlstd{parameter,}\hlkwc{span}\hlstd{=lambda)}
  \hlstd{parameter_grid} \hlkwb{<-} \hlkwd{seq}\hlstd{(}\hlkwd{min}\hlstd{(parameter),} \hlkwd{max}\hlstd{(parameter),} \hlkwc{length.out} \hlstd{= Ngrid)}
  \hlstd{smoothed_loglik} \hlkwb{<-} \hlkwd{predict}\hlstd{(smooth_fit,}\hlkwc{newdata}\hlstd{=parameter_grid)}
  \hlstd{smooth_arg_max} \hlkwb{<-} \hlstd{parameter_grid[}\hlkwd{which.max}\hlstd{(smoothed_loglik)]}
  \hlstd{dist} \hlkwb{<-} \hlkwd{abs}\hlstd{(parameter}\hlopt{-}\hlstd{smooth_arg_max)}
  \hlstd{included} \hlkwb{<-} \hlstd{dist} \hlopt{<} \hlkwd{sort}\hlstd{(dist)[}\hlkwd{trunc}\hlstd{(lambda}\hlopt{*}\hlkwd{length}\hlstd{(dist))]}
  \hlstd{maxdist} \hlkwb{<-} \hlkwd{max}\hlstd{(dist[included])}
  \hlstd{weight} \hlkwb{<-} \hlkwd{rep}\hlstd{(}\hlnum{0}\hlstd{,}\hlkwd{length}\hlstd{(parameter))}
  \hlstd{weight[included]} \hlkwb{<-} \hlstd{(}\hlnum{1}\hlopt{-}\hlstd{(dist[included]}\hlopt{/}\hlstd{maxdist)}\hlopt{^}\hlnum{3}\hlstd{)}\hlopt{^}\hlnum{3}
  \hlstd{quadratic_fit} \hlkwb{<-} \hlkwd{lm}\hlstd{(lp} \hlopt{~} \hlstd{a} \hlopt{+} \hlstd{b,} \hlkwc{weight}\hlstd{=weight,}
    \hlkwc{data} \hlstd{=} \hlkwd{data.frame}\hlstd{(}\hlkwc{lp}\hlstd{=lp,}\hlkwc{b}\hlstd{=parameter,}\hlkwc{a}\hlstd{=}\hlopt{-}\hlstd{parameter}\hlopt{^}\hlnum{2}\hlstd{)}
  \hlstd{)}
  \hlstd{b} \hlkwb{<-} \hlkwd{unname}\hlstd{(}\hlkwd{coef}\hlstd{(quadratic_fit)[}\hlstr{"b"}\hlstd{] )}
  \hlstd{a} \hlkwb{<-} \hlkwd{unname}\hlstd{(}\hlkwd{coef}\hlstd{(quadratic_fit)[}\hlstr{"a"}\hlstd{] )}
  \hlstd{m} \hlkwb{<-} \hlkwd{vcov}\hlstd{(quadratic_fit)}
  \hlstd{var_b} \hlkwb{<-} \hlstd{m[}\hlstr{"b"}\hlstd{,}\hlstr{"b"}\hlstd{]}
  \hlstd{var_a} \hlkwb{<-} \hlstd{m[}\hlstr{"a"}\hlstd{,}\hlstr{"a"}\hlstd{]}
  \hlstd{cov_ab} \hlkwb{<-} \hlstd{m[}\hlstr{"a"}\hlstd{,}\hlstr{"b"}\hlstd{]}
  \hlstd{se_mc_squared} \hlkwb{<-} \hlstd{(}\hlnum{1} \hlopt{/} \hlstd{(}\hlnum{4} \hlopt{*} \hlstd{a}\hlopt{^}\hlnum{2}\hlstd{))} \hlopt{*} \hlstd{(var_b} \hlopt{-} \hlstd{(}\hlnum{2} \hlopt{*} \hlstd{b}\hlopt{/}\hlstd{a)} \hlopt{*} \hlstd{cov_ab} \hlopt{+} \hlstd{(b}\hlopt{^}\hlnum{2} \hlopt{/} \hlstd{a}\hlopt{^}\hlnum{2}\hlstd{)} \hlopt{*} \hlstd{var_a)}
  \hlstd{se_stat_squared} \hlkwb{<-} \hlnum{1}\hlopt{/}\hlstd{(}\hlnum{2}\hlopt{*}\hlstd{a)}
  \hlstd{se_total_squared} \hlkwb{<-} \hlstd{se_mc_squared} \hlopt{+} \hlstd{se_stat_squared}
  \hlstd{delta} \hlkwb{<-} \hlkwd{qchisq}\hlstd{(confidence,}\hlkwc{df}\hlstd{=}\hlnum{1}\hlstd{)} \hlopt{*} \hlstd{( a} \hlopt{*} \hlstd{se_mc_squared} \hlopt{+} \hlnum{0.5}\hlstd{)}
  \hlstd{loglik_diff} \hlkwb{<-} \hlkwd{max}\hlstd{(smoothed_loglik)} \hlopt{-} \hlstd{smoothed_loglik}
  \hlstd{ci} \hlkwb{<-} \hlkwd{range}\hlstd{(parameter_grid[loglik_diff} \hlopt{<} \hlstd{delta])}
  \hlkwd{list}\hlstd{(}\hlkwc{lp}\hlstd{=lp,}\hlkwc{parameter}\hlstd{=parameter,}\hlkwc{confidence}\hlstd{=confidence,}
    \hlkwc{quadratic_fit}\hlstd{=quadratic_fit,} \hlkwc{quadratic_max}\hlstd{=b}\hlopt{/}\hlstd{(}\hlnum{2}\hlopt{*}\hlstd{a),}
    \hlkwc{smooth_fit}\hlstd{=smooth_fit,}
    \hlkwc{fit}\hlstd{=}\hlkwd{data.frame}\hlstd{(}
      \hlkwc{parameter}\hlstd{=parameter_grid,}
      \hlkwc{smoothed}\hlstd{=smoothed_loglik,}
      \hlkwc{quadratic}\hlstd{=}\hlkwd{predict}\hlstd{(quadratic_fit,} \hlkwd{list}\hlstd{(}\hlkwc{b} \hlstd{= parameter_grid,} \hlkwc{a} \hlstd{=} \hlopt{-}\hlstd{parameter_grid}\hlopt{^}\hlnum{2}\hlstd{))}
    \hlstd{),}
    \hlkwc{mle}\hlstd{=smooth_arg_max,} \hlkwc{ci}\hlstd{=ci,} \hlkwc{delta}\hlstd{=delta,}
    \hlkwc{se_stat}\hlstd{=}\hlkwd{sqrt}\hlstd{(se_stat_squared),} \hlkwc{se_mc}\hlstd{=}\hlkwd{sqrt}\hlstd{(se_mc_squared),} \hlkwc{se}\hlstd{=}\hlkwd{sqrt}\hlstd{(se_total_squared)}
  \hlstd{)}
\hlstd{\}}
\end{alltt}
\end{kframe}
\end{knitrout}

\end{document}